# WST, the Wide-field Spectroscopic Telescope: dispersing elements

Andrea Bianco[a,*], Michele Frangiamore[a], Anna Brucalassi[b], Dimitri Buffat[c], Corentin Cudennec[d], Kjetil Dohlen[c], Alexandre Jeanneau[d], David Lee[e], Will Saunders[f], Andrea Tozzi[b]

[a]INAF – Osservatorio Astronomico di Brera, Italy; [b]INAF – Osservatorio Astrofisico di Arcetri, Italy; [c]Laboratoire d'Astrophysique de Marseille, France; [d]Observatoire de Lyon, France; [e]UK Astronomy Technology Centre, United Kingdom; [f]Macquarie University, Australia

## ABSTRACT

The Wide-field Spectroscopic Telescope (WST) is a proposed 12-m class facility entirely dedicated to spectroscopic surveys, combining a high-multiplex multi-object spectrograph operating at low (MOS-LR) and high (MOS-HR) spectral resolution with a giant panoramic integral-field spectrograph (IFS), all three operating in parallel. Diffraction gratings are the key dispersing elements of all three instruments and, given the very large number and size of the units required, drive critical trade-offs in throughput, feasibility and production cost. This paper reviews the two grating technologies under consideration for WST, Volume Phase Holographic Gratings (VPHG) and binary (lithographic, surface-relief) gratings, summarizing their working principles and the parameters that control their diffraction efficiency. We then present the current baseline grating parameters and vendor results for each instrument: low-dispersion, VPHGs for the IFS; a four-arm GRISM layout for MOS-LR; and two competing high-resolution disperser architectures ('8M16D' and '16M4D') for MOS-HR, where binary gratings show a promising path to diffraction efficiencies beyond what is achievable with VPHGs. We conclude with the main open challenges, chiefly the mass production of hundreds of grating units within cost and schedule and the next steps foreseen to consolidate the disperser baseline for WST.

## 1. INTRODUCTION

The Wide-field Spectroscopic Telescope (WST) is a proposed 12-m class, wide-field facility conceived and optimized entirely for spectroscopic surveys, combining a wide field of view (2 deg diameter, 3.1 deg²) with a broad spectral range (0.35–1.6 µm)[1]. The scientific case for WST calls for three instruments operating in parallel behind a single telescope[2]: a low-resolution, high-multiplex multi-object spectrograph (MOS-LR), a high-resolution multi-object spectrograph (MOS-HR), and a giant panoramic integral-field spectrograph (IFS). Table 1 summarizes the top-level requirements of the three instruments.

MOS-LR and MOS-HR are fiber-fed, while the IFS relies on image slicing; MOS and IFS operate in parallel, and Target-of-Opportunity capability is implemented both at telescope and fiber level. In each of the three instruments, the dispersing elements are the optical components that ultimately set the achievable spectral resolution and, through their diffraction efficiency, a major share of the end-to-end system throughput. Because of the large telescope aperture and the high multiplexing required, WST calls for a very large number of gratings: of the order of hundreds depending on the instrument. This combination of performance requirements and production scale makes the choice and procurement of the dispersing elements one of the critical design and risk items for the three instruments. This paper reviews the grating technologies considered for WST and the current baseline and trade-offs for each instrument.

* andrea.bianco@inaf.it, https://wstelescope.eu/

Table 1. WST top-level requirements for the telescope and the three instruments (MOS-LR, MOS-HR, IFS).

| Parameter | Value |
|---|---|
| Telescope aperture | 12 m (100 m²) |
| Telescope field of view | 2 deg diam. (3.1 deg²) |
| Telescope spectral range | 0.35–1.6 µm |
| MOS-LR multiplex | 30 000 |
| MOS-LR resolution | 3 800 @ 420 nm – 4 900 @ 680 nm |
| MOS-LR spectral range | 370–930 nm (simultaneous) |
| MOS-HR multiplex | 2 000 |
| MOS-HR resolution | 40 000 |
| MOS-HR spectral range | 405–440, 455–495, 545–595, 610–670 nm |
| IFS field of view & sampling | 3×3 arcmin², 0.25×0.25 arcsec² spaxel |
| IFS resolution | 4 800 @ 480, 750 nm |
| IFS spectral range | 370–930 nm (simultaneous) |
| IFS patrol field | 13 arcmin diameter |

# 2. DISPERSING ELEMENTS: DESIGN CONSIDERATIONS AND TECHNOLOGIES

## 2.1 General design drivers

WST is a seeing-limited facility, so the size of each spectrograph scales with the telescope diameter $D_T$ for a given resolving power R (eq. 1), where $\chi$ is the angular slit width on sky, m the diffraction order, G the grating line density, W the width of the illuminated grating area and $\lambda$ the wavelength[3].

$$R = \frac{mG\lambda\, W}{\chi\, D_T} \qquad (1)$$

For a fixed resolving power, the main trade-off is between the grating line density G and the size of the collimated beam W that illuminates the grating, which in turn sets the overall size of the spectrograph. A high line density reduces the beam size, but also requires a large angle of incidence (AOI); a large AOI narrows the wavelength range over which the diffraction efficiency (DE) stays high, i.e. it reduces the usable DE bandwidth[4]. The high multiplexing required by MOS-LR and MOS-HR further implies either a very long equivalent entrance slit or multiple spectrograph copies; slicing the entrance slit into several fiber pseudo-slits reduces the effective slit width but correspondingly increases the slit length, or the number of spectrograph units, needed to preserve multiplex. For WST, the disperser trade-off studies have therefore focused primarily on throughput and feasibility, while other properties (transmitted wavefront error, stray-light) are also being tracked.

Concerning the diffraction gratings, different possibilities exist that are more suited depending on the spectrograph design, namely, ruled grating, Volume Phase Holographic Gratings, Binary (lithographic) gratings. In general, they are suited for different angular dispersions (the rate of change of the diffraction angle with respect to a change in wavelength) as shown in Fig.1.

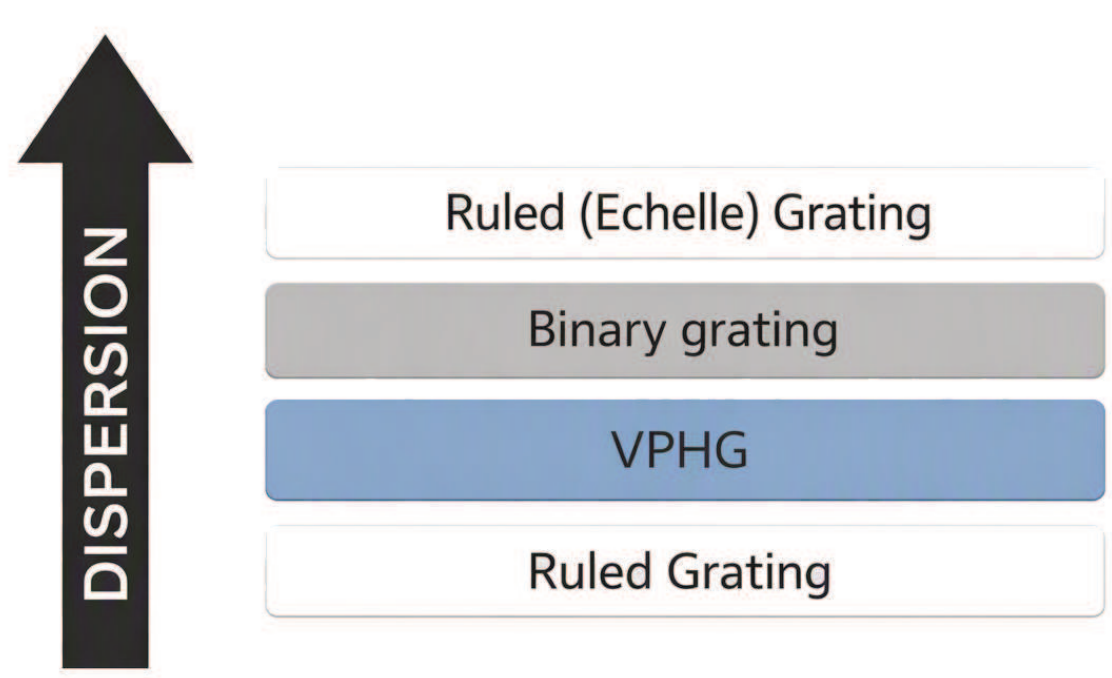

Figure 1. Scheme of the best technology of the diffraction grating as function of the angular dispersion.

In the case of WST with its instruments and target resolving power, we consider the VPHGs and the binary gratings that will be briefly introduced here.

### 2.2 Volume Phase Holographic Gratings

Volume Phase Holographic Gratings (VPHG) record a periodic modulation of the refractive index within a thin layer of dichromated gelatin (DCG) or a photopolymer, sandwiched between two glass substrates[4–9] (Fig. 2 left). Their diffraction efficiency is controlled by the holographic recording material parameters, namely the refractive index modulation Δn and the thickness of the active layer d, which together determine the peak efficiency and its spectral width. A slanted fringe geometry (a fringe tilt relative to the grating surface) is often used to shift the peak of the DE curve away from the Littrow wavelength, which is useful to tailor the response to a specific instrument arm and to avoid Littrow ghosts[10].

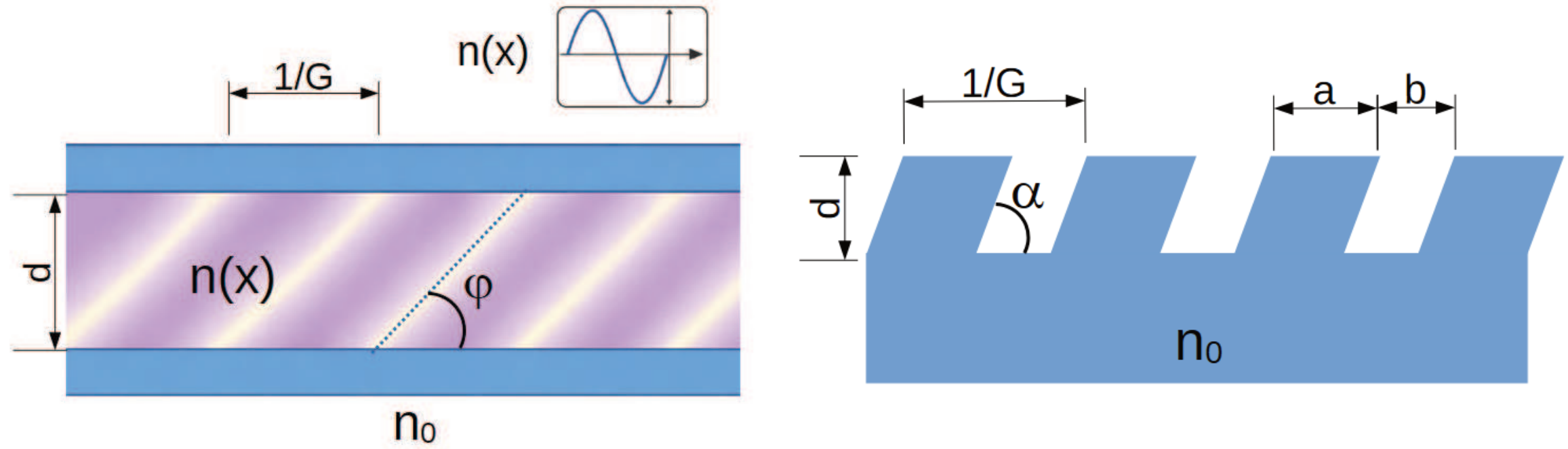

Figure 2. Scheme of a VPHG (left) and a binary grating (right) with the main parameters.

### 2.3 Binary (lithographic) gratings

Binary gratings are produced by lithographic patterning (e.g. laser interference lithography, electron-beam writing[11]) followed by reactive-ion or reactive-ion-beam etching (RIE/RIBE)[12] of a rectangular groove profile (see Fig. 2 right). Commonly they are realized on Fused Silica. Their diffraction efficiency is primarily governed by geometrical parameters, duty cycle and groove depth (thickness). Usually, the inclination angle is difficult to control; moreover, the aspect ratio (b/d) has some limitations because too thin features are difficult to control and to maintain the correct geometry. The grooves can also be backfilled with a material of different refractive index to tune the average index and the index mismatch across the grating, offering an additional degree of freedom to shape the DE curve[12–14].

### 2.4 Companies involved in the analysis

At the time of writing, the parameters of the gratings were determined for the different instruments based on a consolidated design. In order to understand the feasibility, different companies have been inquired. The feasibility was primarily focused to the spectroscopic performances, i.e. the diffraction efficiency (DE) across the target bandwidth. In addition, feedback on the production of large quantities was required. No requirements were set on the wavefront distortion and straylight. Surely, in the continuation of the project, these two aspects will be evaluated. Several companies have been contacted to provide candidate solutions; data have already been received from some vendors, while others are still in progress. Table 2 summarizes the fabrication routes explored for each technology and the typical grating size they can support.

Table 2. Fabrication routes explored for VPHG and binary gratings for WST, and typical achievable grating size.

| Technology | Vendor | Fabrication route | Typical size |
|---|---|---|---|
| VPHG | Wasatch[15] | Holography + dichromated gelatin (DCG) | Large |
| VPHG | INAF | Holography + photopolymer (PP) | Large |
| Binary grating | Horiba[16] | Holography + RIBE/RIE | Large |
| Binary grating | Plymouth gratings[17] | Nanoruler (laser interference lithography) + RIBE/RIE | Large |
| Binary grating | Fraunhofer IOF[18] | E-beam writing + RIBE/RIE/other etching | Medium |

Concerning the VPHGs, Wasatch is on the market for many years, and it uses the technology based on Dichromated Gelatins (DCG). They have the possibility to produce gratings of a size of more than 400 mm. INAF has recently shown the possibility of making VPHGs based on photopolymers and a new production facility is under construction for producing 450 mm gratings[19].
As for the binary grating, HORIBA uses a holographic exposure to impress a photoresist followed by etching to obtain the desired profile. Reflective gratings up to 1 m are produced for high power laser pulse compression. Fused silica transmission gratings can be produced up to 450 mm in diameter. Plymouth grating uses a Scanning Beam Interference Lithography (SBIL) to transfer the optical pattern on the photoresist and then etching techniques to obtain the target pattern in a similar fashion of HORIBA and it produces similar products with similar sizes. IOF employes electron beam lithography in a raster mode to write the periodic pattern on the photoresists and some etching techniques to have the binary pattern. Thanks to the employed ebeam machine, the process is highly versatile. Moreover, they've shown the possibility to tune the diffraction efficiency combining different materials [14]. Concerning the size of the gratings, it is limited to 280 mm because of the ebeam machine.

## 3. DISPERSER SOLUTIONS FOR THE WST INSTRUMENTS

### 3.1 IFS dispersers

The IFS instrument[20] consists in 192 identical spectrographs with two arms: a red arm and a blue arm, covering together the 370–930 nm range. The light is split by a dichroic mirror and the collimated beam is of the order of 200 mm. There are two possible configurations that exploit flat and cylindrical detectors. The main difference is the complexity of the camera; whereas the grating specifications are almost identical. Therefore, we consider here only the flat case. The main parameters are reported in Table 3.

Table 3. Baseline grating parameters for the two IFS spectral arms.

| Spectral arm | AOI (°) | Exit angle (°) | Line density (l/mm) | Spectral coverage (nm) |
|---|---|---|---|---|
| Blue arm | 16.98 | 16.98 | 1182 | 370–595 |
| Red arm | 16.51 | 16.50 | 740 | 575–930 |

The gratings show a low dispersion, with AOIs of ~17°; therefore, the VPHG technology is highly suitable. Moreover, the size is about 200×200 mm² perfectly in the manufacturing range of Wasatch and INAF. The diffraction efficiencies for the two gratings calculated by INAF and Wasatch are reported in Fig. 3.

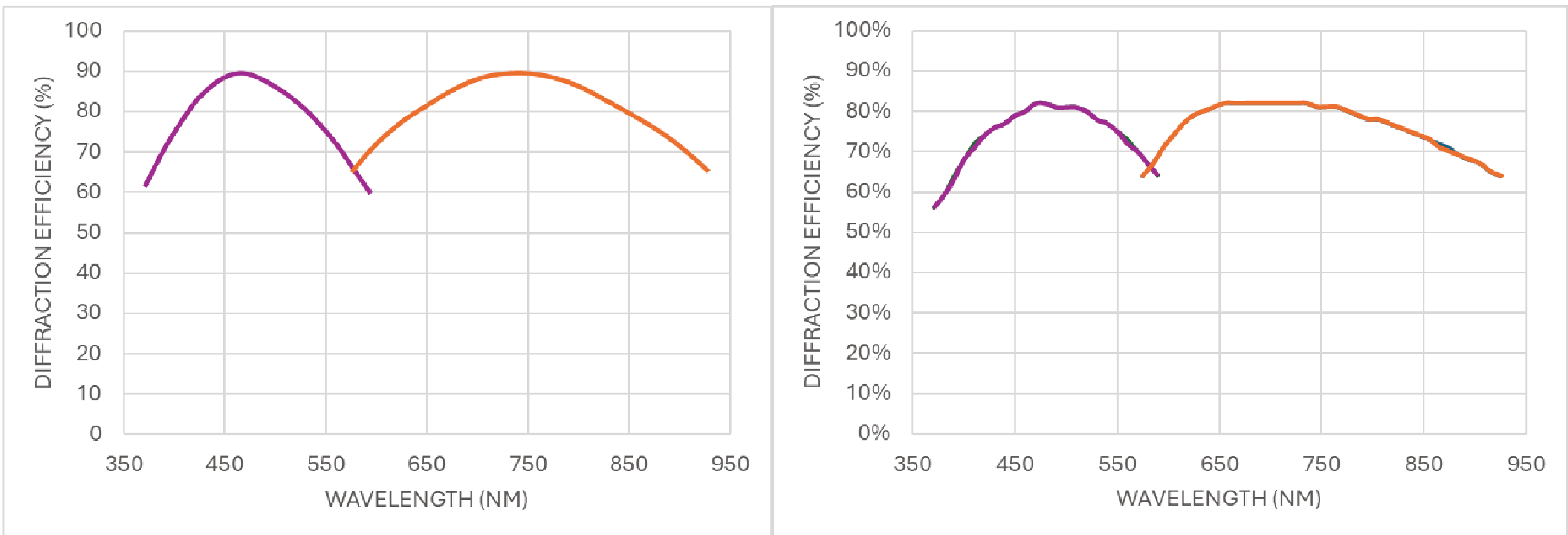

Figure 3. IFS gratings. On the left: Diffraction efficiency curves of the blue and red VPHGs by INAF. On the right: Diffraction efficiency curves by Wasatch.

INAF designed more peaked curves with an average DE of the order of 75-78%. The blue grating shows lower DE at the edges mainly because of residual absorption of the holographic material. Wasatch provided flatter curve with slightly lower DE especially in the bluest part of the spectrum. In general, the performances are very similar. We have, indeed, to consider the fact that the reported curves consider both residual material absorptions and non-idealities of the process. INAF and Wasatch apply different coefficients in this sense, and the DE could differ between the two solutions.

If the feasibility in terms of DE is achieved, indeed, both options are based on mature, low-risk technology, requiring no further R&D. The main point consists in the production process. Indeed, it is necessary to produce 400 items (200 blue and 200 red, considering some spares) in a reasonable amount of time. This also means to procure and store more than 1000 substrates with AR coating that should be broad band to have single type of substrate for both gratings. The target production rate could be of 1 equivalent spectrograph (1 red + 1 blue gratings) per day. Notably, not only is production time consuming, but also the grating characterization. Therefore, it is reasonable to design and build both manufacturing and characterization setups dedicated to the blue and red gratings with costs that are a fraction of the total cost of the items.

### 3.2 MOS-LR dispersers

The MOS-LR spectrograph[21] adopts a four-arm design with GRISM dispersers covering the UB, V, R, and IZ bands. The baseline grating parameters for the four bands are listed in Table 4; the gratings do not operate at the Littrow condition, which is a complication for a binary-grating design. Their Littrow-equivalent AOI is of the order of 30° for the UB, V, R and 36° for the IZ, comparatively larger values than the IFS case, which is necessary to deliver the target resolving power without slicing the on-sky optical fiber.

Table 4. Baseline grating parameters for the four MOS-LR bands (GRISM configuration).

| Band | λmin (nm) | λmax (nm) | Line density (l/mm) | AOI (°) |
|---|---|---|---|---|
| UB | 370 | 484 | 2335 | 24.53 |
| V | 468 | 612 | 1858 | 24.66 |
| R | 594 | 775 | 1471 | 24.73 |
| IZ | 749 | 930 | 1414 | 29.50 |

The collimated beam for the LR spectrographs is slightly larger than the IFS one, which turns into gratings of about 230×220 mm² each. These sizes are easily manufacturable for both VPH and binary grating vendors. To achieve the extremely large multiplexing (30'000), 50-60 spectroscopic units are foreseen that means a total of 240 gratings. Considering the general guidelines reported in Fig. 1, both VPH and binary technologies are suitable. As for the IFS case, the two VPHG vendors (INAF in-house design and Wasatch Photonics) provide a design of LR gratings and the diffraction efficiency curves are shown in Fig. 4.

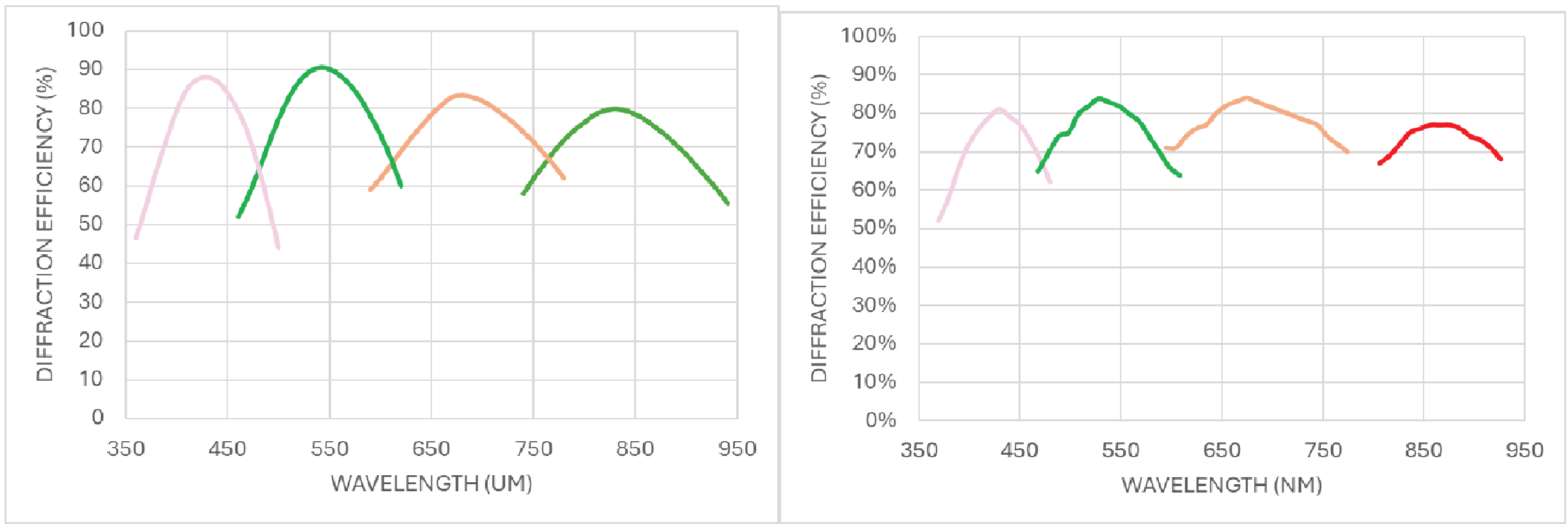

Figure 4. LR gratings. On the left: Diffraction efficiency curves of the four VPHGs by INAF. On the right: Diffraction efficiency curves by Wasatch.

Similarly to the IFS design, the INAF curves are more peaked than the Wasatch curves, with higher peak efficiency and an important drop at the edges, especially in the blue as expected. Despite we have four narrower ranges instead of two wider (IFS), the average DE is slightly smaller (72–77%), showing how the AOI plays a key role.

Plymouth provided a design for the LR gratings and the DE curves are shown in Fig.5. Because for fused silica based binary gratings, the DE is defined by the line density and AOI, the UB, V and R gratings have the same DE curve and only the UB curve is reported together with the IZ one.

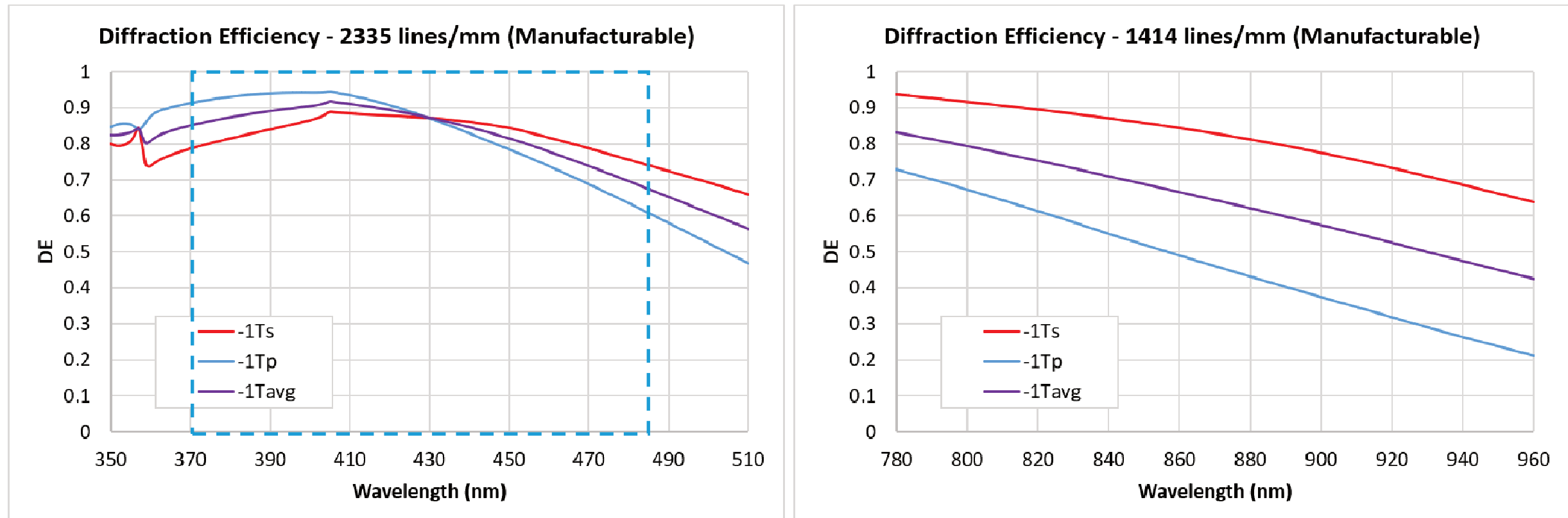

Figure 5. LR gratings, binary gratings by Plymouth. On the left: Diffraction efficiency curves of the UB grating. On the right: Diffraction efficiency curves of the IZ.

For the UB grating, we have a very high DE in short wavelength region (up to 450 nm) and then an important drop, but considering the spectral range, the average DE is 82%. The behavior is the same for the other two bands, V and R. As for the IZ, where the out of Littrow angle is larger, it is evident the effect with a decrease of the DE going from the blue to the red. The average efficiency is significantly slower and the VPH technology behaves better. It is worth noting that these efficiencies are the nominal one and they don't consider possible discrepancies from the theory. In future, it will be necessary to estimate this effect.

Horiba did not provide DE simulations for these LR gratings, but from similar products they could provide similar results of Plymouth. We will expect the complete set of simulations soon.

In conclusion, for the LR instrument, the binary gratings seem to be slightly better than the VPHGs, especially if the design changes in order to reduce the out of Littrow angle. Another option will be to use, the technology that provides the best performances depending on the spectral range.

Concerning the grating production, the numbers are smaller than those of IFS system, but four types are necessary; so, the production should be more than one equivalent spectrograph per week. Moreover, it will not be possible to have only one kind of substrate for the four elements because of the AR coating, increasing the cost of the elements.

### 3.3 MOS-HR dispersers

Two competing architectures are under study for MOS-HR[22]: a 'big' design (8M16D, 8 spectrograph units, gratings 8×4, about 450×660 mm² each) and a 'small' design (16M4D, 16 spectrograph units, gratings 16×4, about 260×290 mm² each), the latter is still under review. The two designs are completely different, the first one is based on very large dispersing elements minimizing the number of spectrographs and keeping the grating AOI to relative small values (45° in air). The latter is much more compact with more spectrographs and reducing importantly the size of the collimated beam and so of the dispersing elements. In this case, the angular dispersion must be higher and the AOI is about 45° in glass, because the first order of diffraction cannot propagate in air. See in table 5 the main parameters of the dispersers.

Table 5. Comparison of the two MOS-HR disperser architectures under study.

| | **8M16D ('big')** | **16M4D ('small')** |
|---|---|---|
| Number of spectrographs | 8 | 16 |
| Grating quantity | 8 × 4 | 16 × 4 |
| Grating size (mm²) | 450 × 660 | 260 × 290 |
| AOI | 45° (in air) | 45°, in glass (GRISM) |
| | | |
| **Spectral ranges** | **Line density (l/mm)** | **Line density (l/mm)** |
| Violet [400.9 - 443.1] | 3315 | 4806 |
| Blue [452.2 - 499.8] | 2971 | 4267 |
| Orange [542.4 - 599.5] | 2476 | 3561 |
| Red [608.0 – 672.0] | 2209 | 3179 |

INAF and Wasatch provided a design for the 8M16D gratings and the corresponding curves are reported in Fig. 6. The results reflect those of the LR gratings with more peaked curves in the INAF case. The average Des are of about 71-74% for INAF and 64-67% for Wasatch. This discrepancy can be due to the different factors used to correct the theoretical curves that are more conservatives for Wasatch.

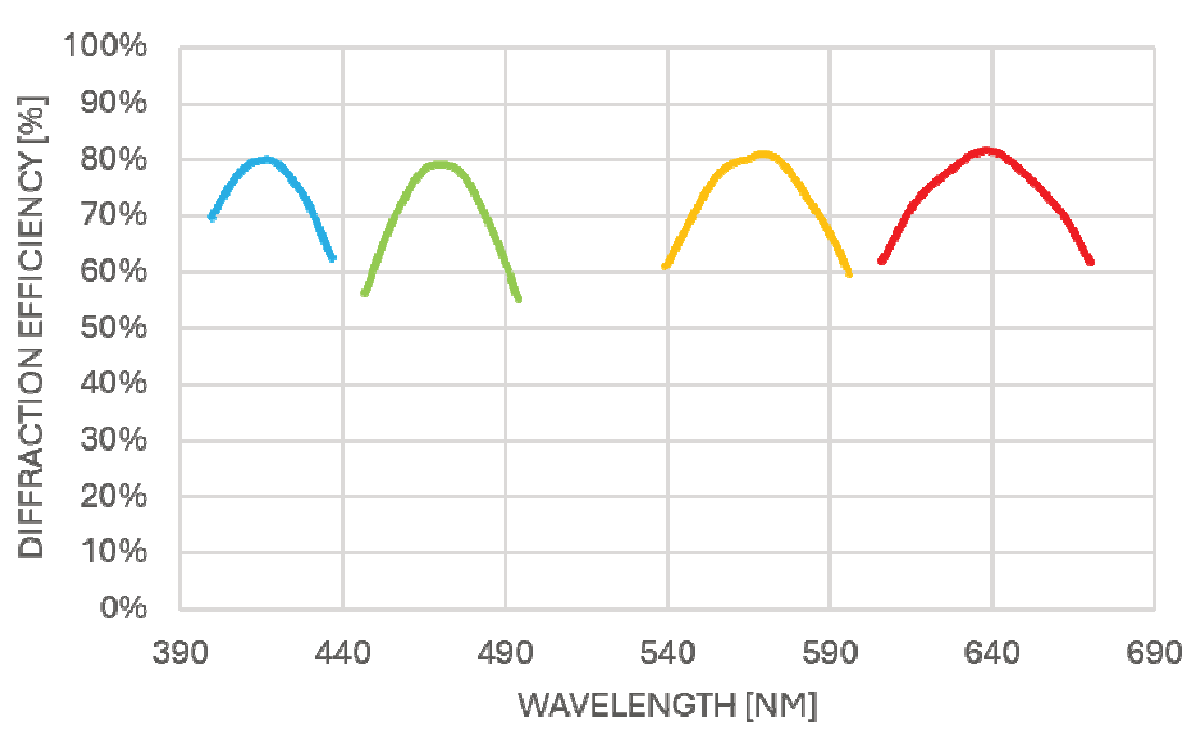

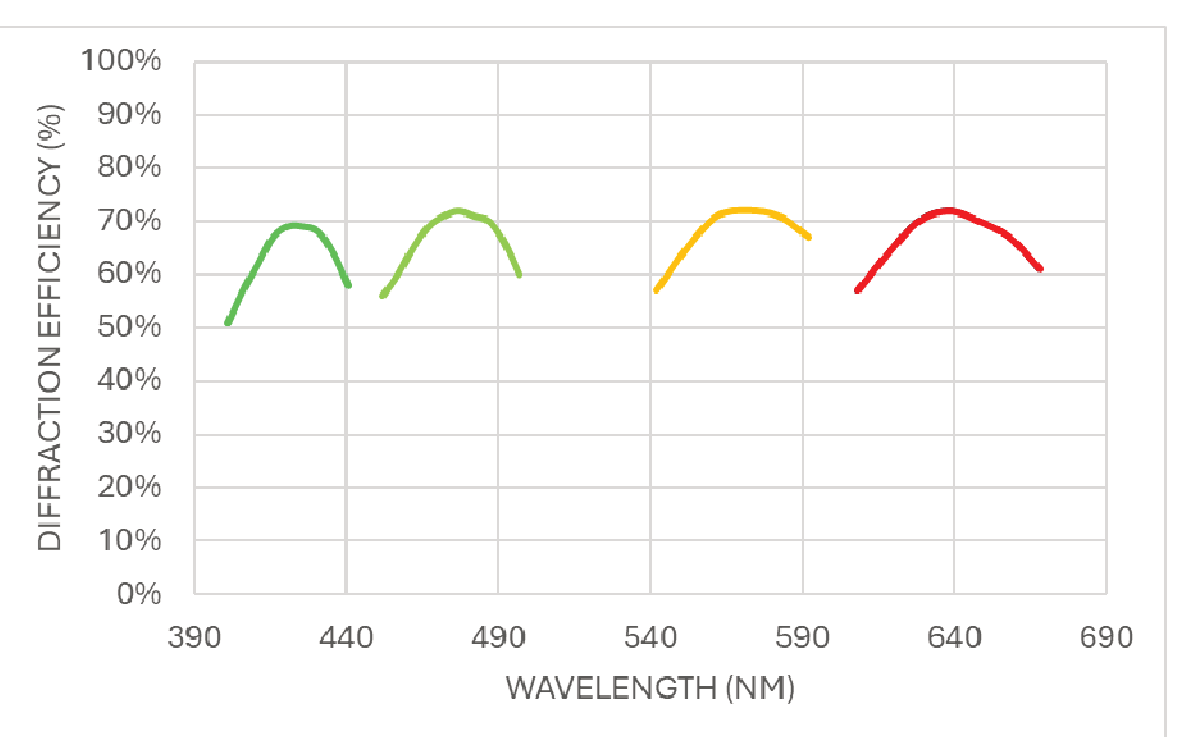

Figure 6. HR VPHGs for 8M16D. On the left: diffraction efficiency curves of the four VPHGs by INAF. On the right: diffraction efficiency curves by Wasatch.

Plymouth calculated the DE curves for their binary gratings, for the 8M16D case and the results are reported in Fig.7 for the blue range. We have to consider that the curves are the same for the other spectral ranges because the AOI remains the same.

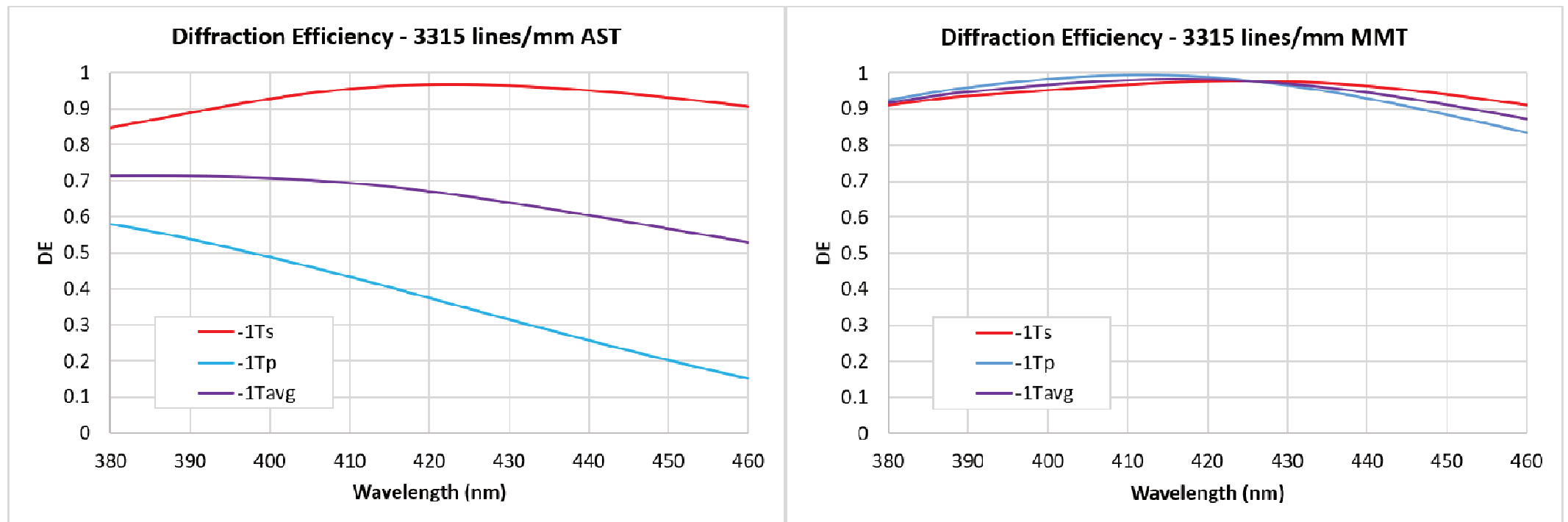

Figure 7. HR binary gratings for 8M16D by Plymouth, blue spectral range. On the left: Standard etching process. On the right: enhanced etching based on R&D activity.

Using the standard process, the DE is limited by the aspect ratio of the binary structures obtaining flat curves with about 65% of average DE mainly because of the large DE mismatch between the two polarizations. It is possible an improvement of the profile based on R&D activity that will make possible to enhance the DE to values of 95% average. In this case, there is no competition with VPH technology and confirm the trend reported in Fig.1, where in high dispersion grating, the binary profile is more effective to provide high DE. Horiba did not provide detailed curves, but only indicative ones showing interesting results. In terms of feasibility, the two vendors can expose the pattern of such size, since they produce meter scale gratings, but an upgrade of the etching facility is necessary because transmission gratings have a limit size of 450 mm in diameter at the moment. Such upgrade is straightforward and it aligns with the WST timeline.

As for the 16M4D design, only Wasatch Photonics provided a design based on VPH technology (see Fig. 8 top). Regarding binary gratings, only IOF designed them.

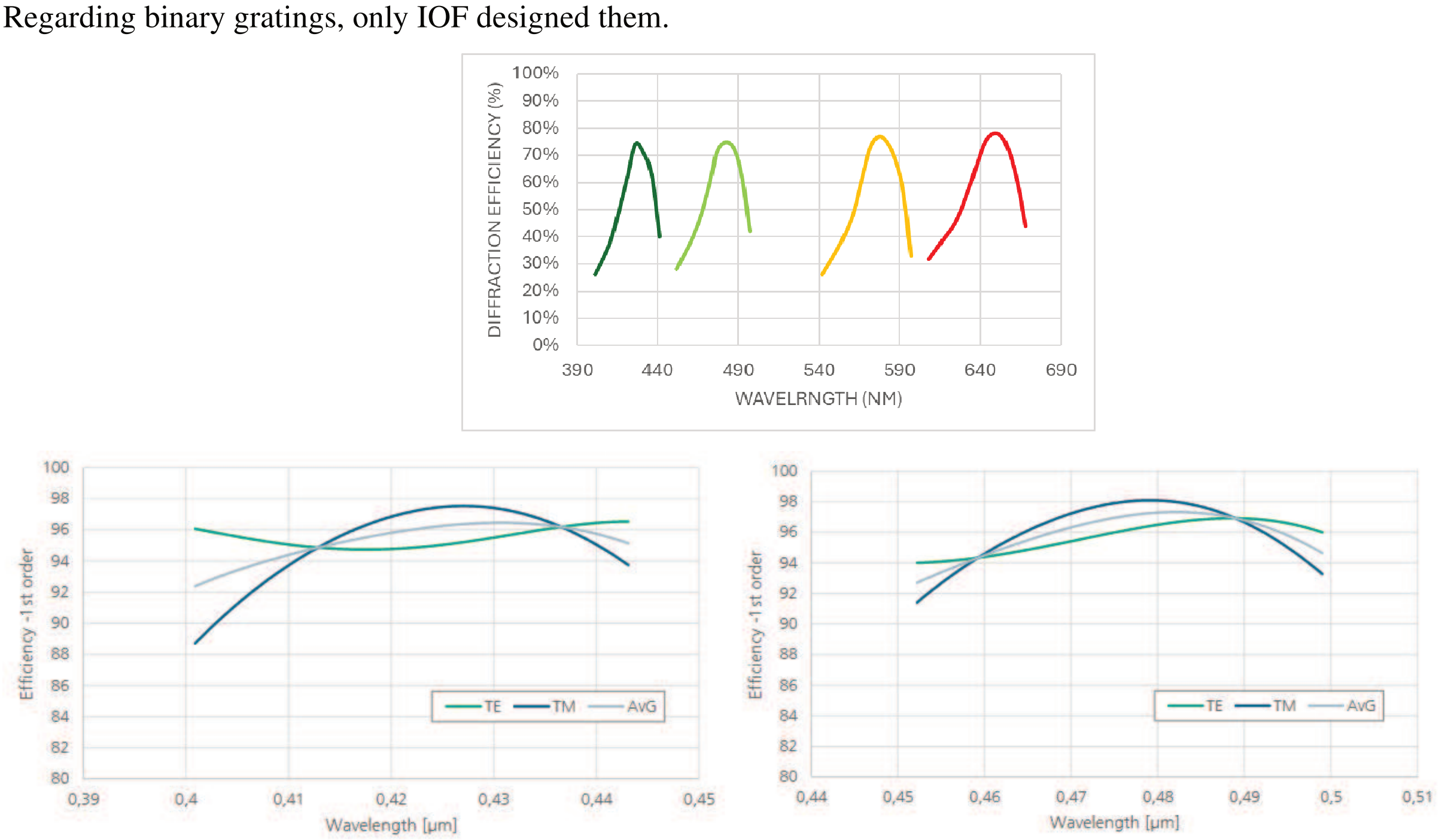

Figure 8. Gratings for 16M4D. On top: DE curves of VPHGs from Wasatch. On bottom: DE curves for binary gratings by IOF for the two bluest ranges.

The VPHGs show highly peaked DE curves with values below 30% at short wavelengths of the spectral ranges. Binary gratings by IOF have very high DEs with average values above 90%. Similar results are expected for the other spectral ranges being the AOIs similar. This is extremely interesting because it increases the total throughput of the HR spectrographs. The size of the elements fits the actual production capabilities of IOF.

## 4. CONCLUSIONS AND PERSPECTIVES

WST would be an ambitious new research infrastructure for the astronomical community providing high multiplexing at low and high resolving power together with a panoramic IFU. The dispersing elements are key components whose requirements differ substantially from instrument to instrument, in size, number, and operating angle of incidence. The trade-off study summarized in this paper indicates that, at present, most of the candidate gratings appear feasible with existing or near-term technology, with no major showstopper identified so far. As a general trend, VPHGs appear best suited to the IFS, binary gratings are the more promising option for MOS-HR, while for MOS-LR both technologies remain possible, and a combination across the four bands is plausible. The main open challenge going forward is how to produce several hundred of gratings, across the three instruments, within a reasonable schedule and cost, while meeting the diffraction-efficiency and feasibility requirements identified for each arm. The consolidation of feasibility, cost and delivery schedule with the candidate vendors is still ongoing. Notably, the companies involved in this study have shown strong interest in the WST project; keeping them engaged through the coming design and procurement phases will be important to secure the disperser baseline for the three instruments.

## ACKNOWLEDGEMENTS

The authors thank the vendors who contributed to the grating design: Zachary Schuberg (Plymouth Grating Laboratory), Elroy Pearson (Wasatch Photonics), Martin Rumpel (Fraunhofer IOF), and William Renard (HORIBA). WST has received funding from the European Union's Horizon Europe research and innovation programme under grant agreement No 101183153.